\documentclass{ws-ijmpa}

\begin{document}

\markboth{A.A. Saharian} {Casimir effect in de Sitter spacetime}

%
\catchline{}{}{}{}{}
%

\title{CASIMIR EFFECT IN DE SITTER SPACETIME}

\author{A. A. SAHARIAN}

\address{Department of Physics, Yerevan State University \\
1 Alex Manoogian Street, 0025 Yerevan, Armenia\\
saharian@ysu.am}

\maketitle

\begin{history}
\received{Day Month Year} \revised{Day Month Year}
\end{history}

\begin{abstract}
The vacuum expectation value of the energy-momentum tensor and the
Casimir forces are investigated for a massive scalar field with an
arbitrary curvature coupling parameter in the geometry of two
parallel plates, on the background of de Sitter spacetime. The
field is prepared in the Bunch--Davies vacuum state and is
constrained to satisfy Robin boundary conditions on the plates.
The vacuum energy-momentum tensor is non-diagonal, with the
off-diagonal component corresponding to the energy flux along the
direction normal to the plates. It is shown that the curvature of
the background spacetime decisively influences the behavior of the
Casimir forces at separations larger than the curvature radius of
de Sitter spacetime. In dependence of the curvature coupling
parameter and the mass of the field, two different regimes are
realized, which exhibit monotonic or oscillatory behavior of the
forces. The decay of the Casimir force at large plate separation
is shown to be power-law, with independence of the value of the
field mass.

\end{abstract}

\ccode{PACS numbers: 04.62.+v, 04.20.Gz, 04.50.-h, 11.10.Kk}

\section{Introduction}

An interesting topic in the investigations of the Casimir effect
(for a review see Refs.~\refcite{Casimir}) is its explicit
dependence on the geometry of background spacetime. Analytic
solutions can be found for highly symmetric geometries only. In
particular, motivated by Randall--Sundrum type braneworld
scenarios, investigations of the Casimir effect in anti-de Sitter
(AdS) spacetime have attracted a great deal of attention. The
braneworld corresponds to a manifold with boundaries and all
fields which propagate in the bulk give Casimir-type contributions
to the vacuum energy and, as a result, to the forces acting on the
branes. The Casimir effect provides in this context a natural
mechanism for stabilizing the radion field.

Another popular background in cosmology is de Sitter (dS)
spacetime. There are several physical motivations for this. dS
spacetime is the maximally symmetric solution of the Einstein
equations with a positive cosmological constant and due to its
high symmetry numerous physical problems are exactly solvable on
this background. A better understanding of physical effects in
this background could serve as a handle to deal with more
complicated geometries. In most inflationary models dS spacetime
is employed to solve a number of problems in standard
cosmology\cite{Lind90}. More recently astronomical observations of
high redshift supernovae, galaxy clusters and cosmic microwave
background\cite{Ries07} indicate that at the present epoch the
universe is accelerating and can be well approximated by a world
with a positive cosmological constant. If the universe would
accelerate indefinitely, the standard cosmology would lead to an
asymptotic dS universe.

In the the present talk, based on Refs.~\refcite{Saha09,Eliz10},
we investigate the vacuum expectation value (VEV) of the
energy-momentum tensor and the Casimir forces for a scalar field
with general curvature coupling parameter in the geometry of two
parallel plates on the background of $(D+1)$-dimensional dS
spacetime (for previous research on the Casimir effect in dS
spacetime see references given in Ref.~\refcite{Saha09}). The
paper is organized as follows. In the next section the VEV of the
energy-momentum tensor is studied for a scalar field with general
curvature coupling parameter and with Robin boundary conditions on
the plates. The Casimir forces acting on the plates are discussed
in section~\ref{sec:Forces}. Section~\ref{sec:Conc} contains a
summary of the work.

\section{VEV of the energy-momentum tensor}

\label{sec:EMT}

We consider a scalar field $\varphi (x)$ with the curvature coupling
parameter $\xi $. The corresponding field equation has the form%
\begin{equation}
\left( \nabla _{l}\nabla ^{l}+m^{2}+\xi R\right) \varphi =0,  \label{fieldeq}
\end{equation}%
where $\nabla _{l}$ is the covariant derivative operator and $R$
is the Ricci scalar for the background spacetime. The values of
the curvature coupling parameter $\xi =0$ and $\xi =\xi _{D}\equiv
(D-1)/4D$, with $D$ being the number of spatial dimensions,
correspond to the most important special cases of minimally and
conformally coupled fields. In the present paper the background
geometry is the $(D+1)$-dimensional dS spacetime. We write the
corresponding line element in planar coordinates, most appropriate
for cosmological applications:%
\begin{equation}
ds^{2}=dt^{2}-e^{2t/\alpha }\sum_{i=1}^{D}(dz^{i})^{2}.  \label{ds2deSit}
\end{equation}%
The Ricci scalar is related to the parameter $\alpha $ in the scale factor
by the formula $R=D(D+1)/\alpha ^{2}$. In addition to the synchronous time
coordinate $t$, we will use the conformal time $\tau $ defined by the
relation $\tau =-\alpha e^{-t/\alpha }$, $-\infty <\tau <0$.

Our main interest are the VEV of the energy-momentum tensor and
the Casimir forces in the geometry of two infinite, parallel
plates located at $z^{D}=a_{j}$,$\;j=1,2$. On the plates the field
obeys Robin boundary conditions (BCs)
\begin{equation}
(1+\beta _{j}n^{l}\nabla _{l})\varphi (x)=0,\quad z^{D}=a_{j},
\label{boundcond}
\end{equation}%
with constant coefficients $\beta _{j}$ and with $n^{l}$ being the normal to
the boundary. For the region between the plates one has $n^{l}=(-1)^{j-1}%
\delta _{D}^{l}$. Dirichlet and Neumann BCs correspond to special cases $%
\beta _{j}=0$ and $\beta _{j}=\infty $, respectively. The
imposition of BCs leads to a modification of the VEVs for physical
quantities, as compared with those in the situation without
boundaries. Among the most important characteristics of the vacuum
state is the VEV of the energy-momentum tensor. In addition to
describing the physical structure of the quantum field at a given
point, the energy-momentum tensor acts as the source in the
Einstein equations and therefore plays an important role in
modelling a self-consistent dynamics involving the gravitational
field. The VEV is
expressed as the mode-sum%
\begin{equation}
\langle 0|T_{ik}|0\rangle =\sum_{\sigma }T_{ik}\{\varphi _{\sigma
}(x),\varphi _{\sigma }^{\ast }(x)\},  \label{EMTVEV}
\end{equation}%
where $\left\{ \varphi _{\sigma }(x),\varphi _{\sigma }^{\ast
}(x)\right\} $ is a complete set of solutions to the classical
field equation satisfying the boundary conditions, and the
bilinear form $T_{ik}\{\varphi (x),\psi (x)\}$ is determined by
the classical energy-momentum tensor. We implicitly assume the
presence of a cutoff function in (\ref{EMTVEV}) which makes the
sum finite.

In the problem under consideration, the time-dependent part of the
eigenfunctions is a linear combination of the functions $\eta ^{D/2}H_{\nu
}^{(l)}(\eta K)$, $l=1,2$, where $H_{\nu }^{(l)}(x)$ is the Hankel function,
$\eta =|\tau |$, and
\begin{equation}
\nu =\left[ D^{2}/4-D(D+1)\xi -m^{2}\alpha ^{2}\right] ^{1/2}.  \label{nu}
\end{equation}%
Different choices of the coefficients in this linear combination
correspond to different choices of the vacuum state. We will
consider dS invariant Bunch--Davies vacuum\cite{Bunc78} for which
the coefficient for the part containing the function $H_{\nu
}^{(2)}(\eta K)$ is zero. In the region between the plates,
$a_{1}<z^{D}<a_{2}$, the corresponding eigenfunctions, satisfying
the BC on the plate at $z^{D}=a_{1}$, have the form
\begin{equation}
\varphi _{\sigma }(x)=C_{\sigma }\eta ^{D/2}H_{\nu }^{(1)}(\eta K)\cos
[k_{D}(z^{D}-a_{1})+\alpha _{1}(k_{D})]e^{i\mathbf{k}\cdot \mathbf{z}},
\label{eigfuncD}
\end{equation}%
with the notations $K=\sqrt{k^{2}+k_{D}^{2}}$ and $e^{2i\alpha
_{1}(x)}=(i\beta _{1}x-1)/(i\beta _{1}x+1)$. In Eq.~(\ref{eigfuncD}), $\;%
\mathbf{z}=(z^{1},\ldots ,z^{D-1})$ is the position vector along the
dimensions parallel to the plates and $\mathbf{k}=(k_{1},\ldots ,k_{D-1})$.
For a conformally coupled massless field one has $\nu =1/2$. From the
boundary condition on the plate $z^{D}=a_{2}$ it follows that the
eigenvalues for $k_{D}$ are solutions of the equation
\begin{equation}
(1-b_{1}b_{2}y^{2})\sin y-(b_{1}+b_{2})y\cos y=0,\;y=k_{D}a,\;
\label{kDvalues}
\end{equation}%
where $b_{j}=\beta _{j}/a$ and $a=a_{2}-a_{1}$. In the discussion
below we will assume that all zeros are real. In particular, this
is the case for the conditions $b_{j}\leqslant 0$ (see
Ref.~\refcite{Rome02}). The positive
solutions of Eq.~(\ref{kDvalues}) will be denoted by $y=\lambda _{n}$, $%
n=1,2,\ldots $, and for the eigenvalues of $k_{D}$ one has $k_{D}=\lambda
_{n}/a$. Consequently, the eigenfunctions are specified by the set $\sigma =(%
\mathbf{k},n)$. The coefficient $C_{\sigma }$ in (\ref{eigfuncD}) is
determined from the normalization condition and is given by the expression%
\begin{equation}
C_{\sigma }^{2}=\frac{(2\pi )^{2-D}\alpha ^{1-D}e^{i(\nu -\nu ^{\ast })\pi
/2}}{4a\left\{ 1+\cos [\lambda _{n}+2\alpha _{1}(\lambda _{n}/a)]\sin
(\lambda _{n})/\lambda _{n}\right\} },  \label{normCD}
\end{equation}%
the star meaning complex conjugate.

It is well known that in dS spacetime without boundaries the Bunch--Davies
vacuum state is not a physically realizable state for {\textrm{Re\thinspace }%
}$\nu \geqslant D/2$. The corresponding two-point functions
contain infrared divergences. In the presence of boundaries, the
BCs on the quantized field may exclude long wavelength modes and
the Bunch--Davies vacuum becomes a realizable state. An example of
this type of situation is provided by the geometry of two parallel
plates described above. In the region between the plates and for
BCs with $\beta _{j}\leqslant 0$, $\beta _{j}\neq \infty $, there
is a maximum wavelength, $2\pi a/\lambda _{1}$, and the two-point
functions contain no infrared divergences. Mathematically, this
situation corresponds to the one where in the argument of the
Hankel function we have $K\geqslant \lambda _{1}/a$.

Combining Eqs.~(\ref{EMTVEV}),(\ref{eigfuncD}),(\ref{normCD}), for the VEV\
of the energy-momentum tensor in the region between the plates we find the
expression which contains series over $\lambda _{n}$. For the summation of
this series we apply the Abel--Plana type summation formula from Refs. %
\refcite{Rome02,Saha08b}. This allows us to write the diagonal components in
the decomposed form (no summation over $l$)
\begin{eqnarray}
\langle T_{l}^{l}\rangle &=&\langle T_{l}^{l}\rangle _{j}+\frac{A_{D}}{%
\alpha ^{D+1}}\int_{0}^{\infty }dy\,y^{1-D}\int_{y}^{\infty }dx\,H(x,y)
\notag \\
&&\times \left[ g(\beta _{j}x/\eta ,|z^{D}-a_{j}|x/\eta
)G_{l}(y)+2G_{l}x^{2}F_{\nu }(y)\right] ,  \label{Tll}
\end{eqnarray}%
where
\begin{eqnarray}
&&A_{D}=\frac{4(4\pi )^{-(D+1)/2}}{\Gamma ((D-1)/2)},\;H(x,y)=\frac{%
(x^{2}-y^{2})^{(D-3)/2}}{c_{1}(x/\eta )c_{2}(x/\eta )e^{2ax/\eta }-1},
\notag \\
&&g(\beta _{j}u,yu)=c_{j}(u)e^{2uy}+e^{-2uy}/c_{j}(u)+2,\;c_{j}(u)=\frac{%
\beta _{j}u-1}{\beta _{j}u+1},  \label{Hxy}
\end{eqnarray}%
and%
\begin{equation}
F_{\nu }(y)=y^{D}\left[ I_{\nu }(y)+I_{-\nu }(y)\right] K_{\nu }(y),
\label{Fnuy}
\end{equation}%
with $I_{\nu }(y)$ and $K_{\nu }(y)$ being the modified Bessel functions. In
(\ref{Tll}), we have introduced the notations%
\begin{eqnarray}
G_{0}(y) &=&\left[ \frac{y^{2}}{4}\partial _{y}^{2}-D(\xi +\xi
_{D})y\partial _{y}+D^{2}\xi +m^{2}\alpha ^{2}-y^{2}+\left( 1-4\xi \right)
x^{2}\right] F_{\nu }(y),  \notag \\
G_{D}(y) &=&\left\{ \left( \xi -\frac{1}{4}\right) y^{2}\partial _{y}^{2}+%
\left[ \xi (2-D)+\frac{D-1}{4}\right] y\partial _{y}-\xi D\right\} F_{\nu
}(y),  \label{Fhat} \\
G_{l}(y) &=&G_{D}(y)+\left[ \frac{y^{2}-x^{2}}{D-1}+\left( 1-4\xi \right)
x^{2}\right] F_{\nu }(y),\;l=1,\ldots ,D-1,  \notag
\end{eqnarray}%
and $G_{D}=1$,\ $G_{l}=4\xi -1$ for\ $l=0,1,\ldots ,D-1$. In Eq.~(\ref{Tll})
(no summation over $l$),%
\begin{equation}
\langle T_{l}^{l}\rangle _{j}=\langle T_{l}^{l}\rangle _{\text{dS}}+\frac{%
A_{D}}{\alpha ^{D+1}}\int_{0}^{\infty }dy\,y^{1-D}\int_{y}^{\infty
}dx\,(x^{2}-y^{2})^{\frac{D-3}{2}}\frac{e^{-2x|z^{D}-a_{j}|/\eta }}{%
c_{j}(x/\eta )}G_{l}(y),  \label{Tll1j}
\end{equation}%
is the VEV for the geometry of a single plate at $z^{D}=a_{j}$
when the second plate is absent\cite{Saha09} and $\langle
T_{l}^{l}\rangle _{\text{dS}}$ is the corresponding renormalized
VEV in dS spacetime without boundaries. For points away from the
plates, renormalization is required for the latter part only. Due
to the dS invariance of the Bunch--Davies vacuum, $\langle
T_{k}^{l}\rangle _{\text{dS}}$ is proportional to the metric
tensor with a
constant coefficient and has been well investigated in the literature\cite%
{Bunc78,Cand75}. The last term on the right hand side of Eq.~(\ref{Tll}) is
induced by the presence of the second plate. Note that in the formulas given
above, $|z^{D}-a_{j}|/\eta $ is the proper distance of the observation point
from the plate at $z^{D}=a_{j}$, measured in units of the dS curvature
radius $\alpha $. The VEVs depend on time through the combinations $%
|z^{D}-a_{j}|/\eta $ and $\beta _{j}/\eta $. This property is a
consequence of the maximal symmetry of dS spacetime and of
Bunch--Davies vacuum.

For the non-zero off-diagonal component, we have%
\begin{eqnarray}
\langle T_{0}^{D}\rangle &=&\langle T_{0}^{D}\rangle _{j}-\text{sgn}%
(z^{D}-a_{j})\frac{A_{D}}{2\alpha ^{D+1}}\int_{0}^{\infty }dyy^{1-D}G_{0D}(y)
\notag \\
&&\times \int_{y}^{\infty }dx\,xH(x,y)\left[ c_{j}(x/\eta
)e^{2x|z^{D}-a_{j}|/\eta }-e^{-2x|z^{D}-a_{j}|/\eta }/c_{j}(x/\eta )\right] ,
\label{T0Dn}
\end{eqnarray}%
where the part corresponding to the geometry of a single plate is given by%
\begin{eqnarray}
\langle T_{0}^{D}\rangle _{j} &=&\text{sgn}(z^{D}-a_{j})\frac{2A_{D}}{\alpha
^{D+1}}\int_{0}^{\infty }dyy^{1-D}G_{0D}(y)  \notag \\
&&\times \int_{y}^{\infty }dx\,x(x^{2}-y^{2})^{\frac{D-3}{2}}\frac{%
e^{-2x|z^{D}-a_{j}|/\eta }}{c_{j}(x/\eta )}.  \label{T0Dj}
\end{eqnarray}%
In these formulas we have defined the function%
\begin{equation}
G_{0D}(y)=\left[ (4\xi -1)y\partial _{y}+4\xi \right] F_{\nu }(y).
\label{F0D}
\end{equation}%
The off-diagonal component (\ref{T0Dn}) corresponds to the energy flux along
the direction perpendicular to the plates. Depending on the values of the
coefficients in the boundary conditions and of the field mass this flux can
be positive or negative. In the case when $\beta _{1}=\beta _{2}$, the
off-diagonal component $\langle T_{0}^{D}\rangle $ vanishes at $%
z^{D}=(a_{1}+a_{2})/2$. This property is a direct consequence of the problem
symmetry.

For a conformally coupled massless scalar field ($\xi =\xi _{D}$, $m=0$) the
single plate part in the VEV of the energy-momentum tensor vanishes and one
finds (no summation over $l$)
\begin{equation}
\langle T_{k}^{l}\rangle =\langle T_{k}^{l}\rangle _{\text{dS}}-\frac{(\eta
/\alpha )^{D+1}B_{l}\delta _{l}^{k}}{(4\pi )^{D/2}\Gamma (D/2+1)}\
\int_{0}^{\infty }dx\,\frac{x^{D}}{c_{1}(x)c_{2}(x)e^{2ax}-1}\,,
\label{TllConf}
\end{equation}%
where $B_{l}=1$ for $\ l=0,\ldots ,D-1$ and $B_{D}=-D$. The boundary induced
part in this formula could have been obtained from the corresponding result
for the Casimir effect in Minkowski spacetime, by using the fact that the
two problems are conformally related (see also Refs. %
\refcite{Saha03,Saha10RW} for the Casimir densities in AdS and
Robertson-Walker spacetimes). Note that the boundary induced part in Eq.~(\ref%
{TllConf}) is traceless and the trace anomaly is contained in the
boundary-free part only.

In the region $z^{D}<a_{1}$ ($z^{D}>a_{2}$) the VEV of the energy-momentum
tensor coincides with the corresponding VEV for a single plate located at $%
z^{D}=a_{1}$ ($z^{D}=a_{2}$) and is given by the expressions (\ref{Tll1j})
and (\ref{T0Dj}), with $j=1$ ($j=2$). The results obtained in the present
paper can be applied to a more general problem where the cosmological
constant is different in separate regions $z^{D}<a_{1}$, $a_{1}<z^{D}<a_{2}$%
, and $z^{D}>a_{2}$. In this case the plate can be considered as a simple
model of a thin domain wall separating the regions with different dS vacua.

In the discussion above we have considered the VEV of the bulk
energy-momentum tensor. For scalar fields with general curvature coupling
parameter and with Robin BCs, in Ref. \refcite{Rome02} it has been shown
that in the discussion of the relation between the mode sum energy and the
volume integral of the renormalized energy density for the Robin parallel
plates geometry in Minkowski spacetime it is necessary to include in the
energy a surface term concentrated on the boundary. An expression for the
surface energy-momentum tensor for a scalar field with a general curvature
coupling parameter in the general case of bulk and boundary geometries is
derived in Ref. \refcite{SahaEMT}. The investigation of the VEV for the
surface densities in the problem under consideration will be reported in
Ref. \refcite{SahaSurf}.

\section{Casimir forces}

\label{sec:Forces}

Having the VEV of the energy-momentum tensor, we can evaluate the forces
acting on the plates. The vacuum force acting per unit surface of the plate
at $z^{D}=a_{j}$ is determined by the $_{D}^{D}$-component of the vacuum
energy-momentum tensor evaluated at this point. For the region between the
plates, the corresponding effective pressures can be written as $%
p^{(j)}=p_{1}^{(j)}+p_{\text{(int)}}^{(j)}$, $j=1,2$. The term $p_{1}^{(j)}$
is the pressure for a single plate at $z^{D}=a_{j}$, when the second plate
is absent. This term is divergent due to the surface divergences in the
subtracted VEVs and needs additional renormalization. The term $p_{\text{%
(int)}}^{(j)}$ is the pressure induced by the second plate, and can be
termed as an interaction force. This contribution is finite for all nonzero
distances between the plates. In the regions $z^{D}<a_{1}$ and $z^{D}>a_{2}$
we have $p^{(j)}=p_{1}^{(j)}$. As a result, the contributions to the vacuum
force coming from the term $p_{1}^{(j)}$ are the same from the left and from
the right sides of the plate, so that there is no net contribution to the
effective force.

The interaction force on the plate at $z^{D}=a_{j}$ is obtained
from the last term on the right hand side of Eq.~(\ref{Tll}) for
$\langle T_{D}^{D}\rangle $
(with minus sign) taking $z^{D}=a_{j}$:%
\begin{equation}
p_{\text{(int)}}^{(j)}=-\frac{2A_{D}}{\alpha ^{D+1}}\int_{0}^{\infty
}dy\,y^{1-D}\int_{y}^{\infty }dx\,x^{2}H(x,y)\left[ \frac{2\left( \beta
_{j}/\eta \right) ^{2}G_{D}(y)}{\left( \beta _{j}x/\eta \right) ^{2}-1}%
+F_{\nu }(y)\right] ,  \label{pintj}
\end{equation}%
where $H(x,y)$ is defined by Eq. (\ref{Hxy}). The time dependence of the
forces appears in the form $a/\eta $ and $\beta _{j}/\eta $. Note that the
ratio $a/\eta $ is the proper distance between the plates measured in units
of dS curvature radius $\alpha $. The effective pressures (\ref{pintj}) can
be either positive or negative, leading to repulsive or to attractive
forces, respectively. For $\beta _{1}\neq \beta _{2}$ the Casimir forces
acting on the left and on the right plates are different. For large values
of $\alpha $, to leading order, the corresponding result for the geometry of
two parallel plates in Minkowski spacetime is obtained:
\begin{equation}
p_{\text{(int)}}^{(j)}\approx p_{\text{(M)}}^{(j)}=-\frac{2(4\pi )^{-D/2}}{%
\Gamma (D/2)}\int_{m}^{\infty }dx\,\,\frac{x^{2}(x^{2}-m^{2})^{D/2-1}}{%
c_{1}(x)c_{2}(x)e^{2ax}-1}.  \label{pjMink}
\end{equation}%
Note that in the Minkowski spacetime the force is the same for both plates
with independence of the values for the coefficients $\beta _{j}$ and this
force does not depend on the curvature coupling parameter (for the
interaction forces between the Robin plates in the geometry with an
arbitrary internal space see Ref. \refcite{Eliz09}).

In the special cases of Dirichlet and of Neumann boundary conditions one
finds:%
\begin{eqnarray}
p_{\text{(int)}}^{(\text{D})} &=&-\frac{4\alpha ^{-D-1}}{(2\pi )^{\frac{D}{2}%
+1}}\sum_{n=1}^{\infty }\int_{0}^{\infty }dy\,yF_{\nu }(y)\left[ (D-1)f_{%
\frac{D}{2}}(yu_{n})+f_{\frac{D}{2}-1}(yu_{n})\right] ,  \label{pjD} \\
p_{\text{(int)}}^{(\text{N})} &=&p_{\text{(int)}}^{(\text{D})}-\frac{8\alpha
^{-D-1}}{(2\pi )^{\frac{D}{2}+1}}\sum_{n=1}^{\infty }\int_{0}^{\infty }dy\,%
\frac{G_{D}(y)}{y}f_{\frac{D}{2}-1}(yu_{n}),\;u_{n}=2na/\eta ,  \label{pjN}
\end{eqnarray}%
where $f_{\mu }(x)=K_{\mu }(x)/x^{\mu }$. For $0\leqslant \nu <1$ the
integrand in the expression for $p_{\text{(int)}}^{(\text{D})}$ is positive
which corresponds to an attractive force for all separations.

Now we turn to the investigation of the asymptotic behavior for the vacuum
forces in the general case of Robin BC. In the limit of small proper
distances between the plates, $a/\eta \ll 1$, to leading order we find%
\begin{equation}
p_{\text{(int)}}^{(j)}\approx -\frac{2(\eta /\alpha )^{D+1}}{(4\pi
)^{D/2}\Gamma (D/2)}\int_{0}^{\infty }dx\,\frac{x^{D}}{%
c_{1}(x)c_{2}(x)e^{2ax}-1}.  \label{pintjsmall}
\end{equation}%
If, in addition, $|\beta _{j}|/a\gg 1$, one has%
\begin{equation}
p_{\text{(int)}}^{(j)}\approx -\frac{D\Gamma ((D+1)/2)\zeta _{\text{R}}(D+1)%
}{(4\pi )^{(D+1)/2}(\alpha a/\eta )^{D+1}},  \label{pintjsmall2}
\end{equation}%
and the corresponding force is attractive. In (\ref{pintjsmall2}), $\zeta _{%
\text{R}}(x)$ is the Riemann zeta function. The same result is obtained for
Dirichlet BCs on both plates. In the case of Dirichlet BC on one plate and
non-Dirichlet one on the other, the leading term is obtained from Eq.~(\ref%
{pintjsmall2}) with an additional factor $(2^{-D}-1)$. In this case the
vacuum force is repulsive at small distances.

In considering the large distance asymptotics, corresponding to $a/\eta \gg
1 $, the cases of real and imaginary $\nu $ must be studied separately. For
positive values of $\nu $, one has%
\begin{equation}
p_{\text{(int)}}^{(j)}\approx -\frac{2\alpha ^{-D-1}g_{\nu }^{(j)}\Gamma
(\nu )}{\pi ^{D/2+1}(2a/\eta )^{D-2\nu +2}},  \label{pjlargeNN}
\end{equation}%
for non-Neumann BCs on the plate at $z^{D}=a_{j}$ ($|\beta _{j}|<\infty $)
and
\begin{equation}
p_{\text{(int)}}^{(j)}\approx -\frac{\alpha ^{-D-1}g_{\nu }^{\text{N}%
(j)}\Gamma (\nu )}{\pi ^{D/2+1}(2a/\eta )^{D-2\nu }},  \label{pjlarge}
\end{equation}%
for Neumann BC ($\beta _{j}=\infty $). Here the notations are as follows:%
\begin{eqnarray}
g_{\nu }^{(j)} &=&\left( \frac{D+1}{2}-\nu \right) \Gamma (D/2-\nu +1)\left[
1-2\left( \frac{\beta _{j}}{\eta }\right) ^{2}f_{D}\right]
\sum_{n=1}^{\infty }\frac{(\delta _{1}\delta _{2})^{n}}{n^{D-2\nu +2}},
\label{gnuj} \\
g_{\nu }^{\text{N}(j)} &=&\Gamma (D/2-\nu )f_{D}\sum_{n=1}^{\infty }\frac{%
(\delta _{1}\delta _{2})^{n}}{n^{D-2\nu }},  \notag
\end{eqnarray}%
with $f_{D}=-2\nu \left[ \xi +\left( \xi -1/4\right) (D-2\nu
)\right] $ and $\delta _{j}=c_{j}(0)$. Note that $\delta _{j}=-1$
for non-Neumann BC, while $\delta _{j}=1$ if the BC is Neumann. In
the case of non-Neumann BCs we have assumed that $|\beta
_{j}|/a\ll 1$.

As it is seen from (\ref{pjlarge}), for positive values of $\nu $ and when $%
f_{D}\neq 0$, at large distances the ratio of the Casimir forces acting on
the plate with Neumann and non-Neumann BCs is of the order $(a/\eta )^{2}$.
Note that in neither of these cases does the force depend on the specific
value of Robin coefficient in the BC on the second plate. For Dirichlet BC
on the plate at $z^{D}=a_{j}$ ($\beta _{j}=0$), at large separations the
Casimir force acting on that plate is repulsive (attractive) for Neumann
(non-Neumann) BCs on the other plate. The nature of the force acting on the
plate with Neumann BC depends on the sign of $f_{D}$ and can be either
repulsive or attractive, in function of the curvature coupling parameter and
of the field mass. For minimally and conformally coupled massive scalar
fields one has $f_{D}=\nu (D/2-\nu )$ and $f_{D}=\nu (1/2-\nu )/D$,
respectively, and this parameter is positive. The corresponding force is
attractive (repulsive) for Neumann (non-Neumann) BC on the second plate.
Note that for the geometry of parallel plates in the Minkowski bulk the
Casimir forces at large distances are repulsive for Neumann BC on one plate
and for non-Neumann BC on the other plate. For all other cases of BCs the
forces are attractive.

For imaginary $\nu $, the leading order terms at large separations between
the plates are in the form%
\begin{eqnarray}
p_{\text{(int)}}^{(j)} &\approx &-\frac{4\alpha ^{-D-1}|g_{\nu }^{(j)}|}{\pi
^{D/2+1}(2a/\eta )^{D+2}}\cos [2|\nu |\ln (2a/\eta )+\phi _{(j)}],\;|\beta
_{j}|<\infty ,  \notag \\
p_{\text{(int)}}^{(j)} &\approx &-\frac{2\alpha ^{-D-1}|g_{\nu }^{\text{N}%
(j)}|}{\pi ^{D/2+1}(2a/\eta )^{D}}\cos [2|\nu |\ln (2a/\eta )+\phi _{(j)}^{%
\text{N}}],\;\beta _{j}=\infty ,  \label{pjlargeim}
\end{eqnarray}%
where the phases are defined in accordance with $g_{\nu }^{(j)}=|g_{\nu
}^{(j)}|e^{i\phi _{(j)}}$ and $g_{\nu }^{\text{N}(j)}=|g_{\nu }^{\text{N}%
(j)}|e^{i\phi _{(j)}^{\text{N}}}$. In this case the decay of the vacuum
forces is oscillatory.

Having in mind that spectral properties of spin 2, 1, 0 operators
for dS spacetime are known, the current study can be extended to
the calculation of the Casimir force due to quantum gravity (for
the one-loop effective action of arbitrary quantum gravity in dS
spacetime see Ref.~\refcite{Cogn05}). Note that the calculations
can be extended to a self-interacting scalar field theory too, and
the results described here can be used to study the
curvature-induced phase transitions of the in-in effective
potential in the same way as it was proposed for the out-in
effective potential in Ref.~\refcite{Buch85}.

\begin{figure}[tbph]
\begin{center}
\begin{tabular}{cc}
\epsfig{figure=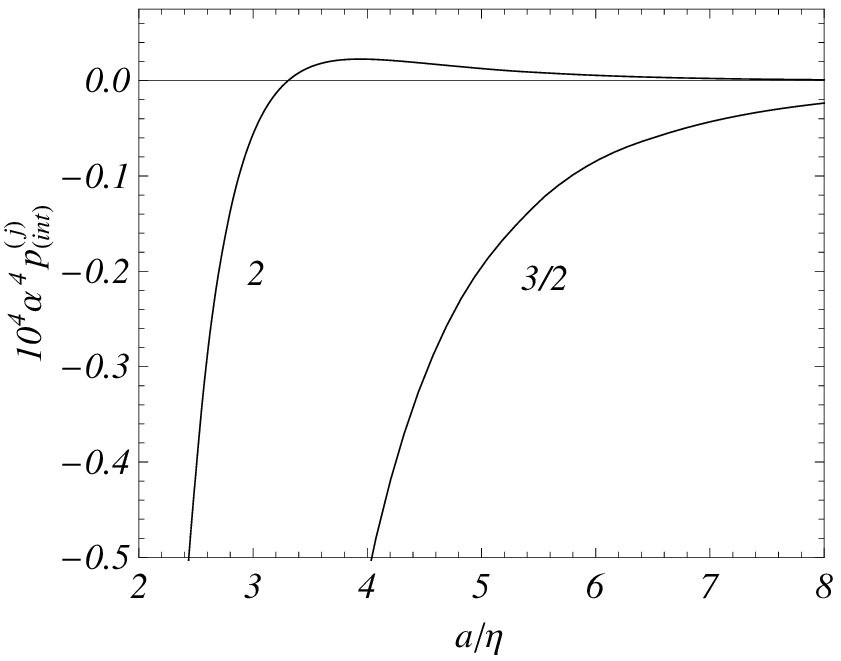,width=5.5cm,height=4.5cm} & \quad %
\epsfig{figure=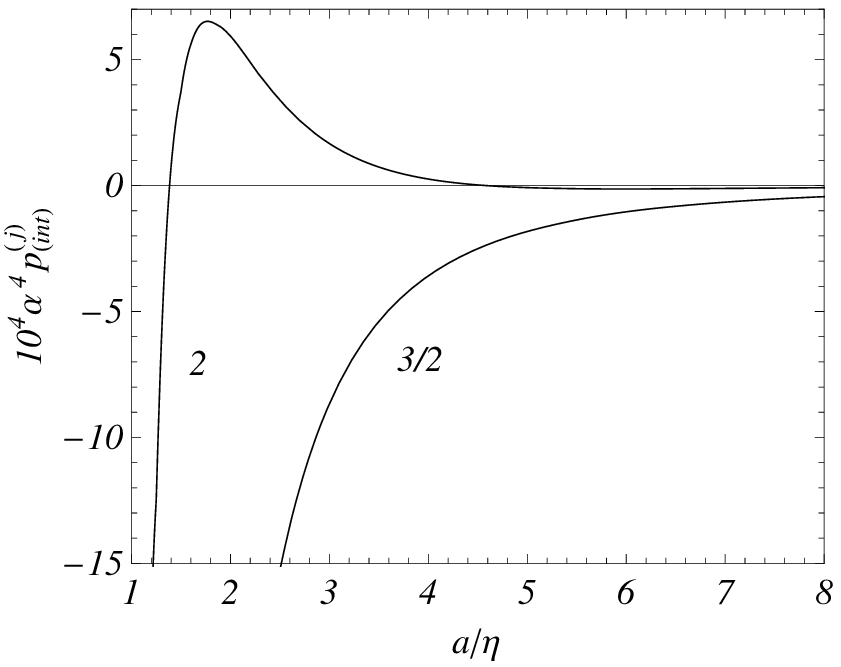,width=5.5cm,height=4.5cm}%
\end{tabular}%
\end{center}
\caption{Interaction forces between the plates for a $D=3$
minimally coupled scalar field with Dirichlet (left plot) and
Neumann (right plot) BCs. The figures near the curves are the
values of the parameter $\protect\alpha m$.} \label{fig1}
\end{figure}

In Fig.~\ref{fig1}, we have plotted the Casimir force for a $D=3$
scalar field with Dirichlet BC, minimally coupled to gravity, as a
function of the proper distance between the plates, measured in
units of the dS curvature scale $\alpha $. The figures near the
curves correspond to the values of the parameter $m\alpha $.
Values are taken in a way so to have both possibilities, with
positive and purely imaginary values of the parameter $\nu $.

\begin{figure}[tbph]
\begin{center}
\epsfig{figure=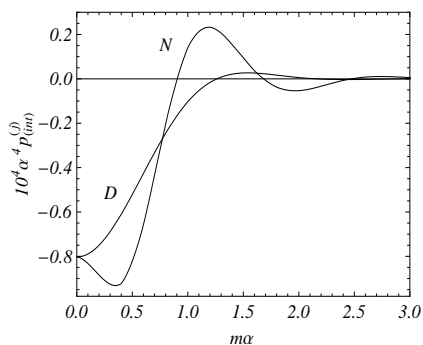,width=5.5cm,height=4.5cm}
\end{center}
\caption{Interaction force between the plates for $a/\protect\eta
=4$ as a function of the field mass, for a $D=3$ conformally
coupled scalar field with Dirichlet and Neumann BCs. }
\label{fig2}
\end{figure}

In Fig.~\ref{fig2} the dependence of the Casimir force on the parameter $%
m\alpha $ is depicted for a given separation corresponding to
$a/\eta =4$. Conformally coupled scalar fields with Dirichlet and
Neumann BCs are considered. For a massless field the force is the
same for Dirichlet and Neumann BCs.

From the discussion given above it follows that for proper
distances between the plates larger than the curvature radius of
the dS spacetime, $\alpha a/\eta \gtrsim \alpha $, the
gravitational field essentially changes the behavior of the
Casimir forces compared with the case of the plates in Minkowski
spacetime. The forces may become repulsive at large separations
between the plates. In particular, for real values $\nu $ and for
Neumann BC on both plates, Casimir forces are repulsive at large
separations, in the range of parameters for which $f_{D}<0$.
Recall that, for the geometry of parallel plates on the background
of Minkowski spacetime, the only case with repulsive Casimir
forces at large distances corresponds to Neumann BC on one plate
and non-Neumann BC on the other. A remarkable feature of the
influence of the gravitational field is the oscillatory behavior
of the Casimir forces at large distances, which appears in the
case of imaginary $\nu $. In this case, the values of the plate
distance yielding zero Casimir force correspond to equilibrium
positions. Among them, the positions with negative derivative of
the force with respect to the distance are locally stable. As
it follows from asymptotic formulas (\ref{pjlargeNN}), (\ref{pjlarge}), and (%
\ref{pjlargeim}), at large separations between the plates the decay of the
Casimir forces as functions of the distance is power-law for both cases of
massive and massless fields. Recall that, in Minkowski spacetime the
corresponding Casimir forces decay as $1/a^{D+1}$ for a massless field and
they are exponentially suppressed by the factor $\exp (-2ma)$ for a massive
filed.

It is also of interest to compare the features for the Casimir
force in dS spacetime with the behavior of the Casimir forces for
parallel plates in AdS spacetime. In Poincar\'{e} coordinates the
corresponding line element is given by the expression
\begin{equation}
ds_{\text{AdS}}^{2}=e^{-\lambda y}\eta _{ik}dx^{i}dx^{k}-dy^{2},
\label{dsAds}
\end{equation}%
where $\eta _{ik}=$diag$(1,-1,\ldots ,-1)$ is the metric tensor for $D$%
-dimensional Minkowski spacetime. For the corresponding Ricci scalar one has
$R=-D(D+1)\lambda ^{2}$ and the AdS curvature radius is given by $1/\lambda $%
. For the general case of Robin BCs on two parallel plates, located at $%
y=y_{j}$, $j=1,2$, the interaction forces between the plates are
investigated in Ref.~\refcite{Saha05} (see also
Refs.~\refcite{Saha06} for the case where an extra compact
subspace is present). At large distances between the plates, as
compared with the AdS curvature radius, $\lambda
(y_{2}-y_{1})\gg 1$, the vacuum interaction forces per unit surface, $p_{%
\text{(int)}}^{(j)}$, are exponentially suppressed by the factor
$\exp [2\nu _{\text{AdS}}\lambda (y_{1}-y_{2})]$ for the plate at
$y=y_{1}$ and by the factor $\exp [(2\nu _{\text{AdS}}+D)\lambda
(y_{1}-y_{2})]$ for the plate at $y=y_{2}$, where $\nu
_{\text{AdS}}=\left[ D^{2}/4-D(D+1)\xi +m^{2}/\lambda ^{2}\right]
^{1/2}$. Note that in AdS spacetime the ground state becomes
unstable for imaginary values of $\nu _{\text{AdS}}$\cite{Brei82}.
Hence, in AdS spacetime the Casimir forces are exponentially
suppressed for both massive and massless fields.

\section{Conclusion}

\label{sec:Conc}

The natural appearance of dS spacetime in a variety of situations
has stimulated considerable interest in the behavior of quantum
fields propagating in this background. In the present paper we
have studied the VEV of the energy--momentum tensor and the
Casimir forces for a scalar field with an arbitrary curvature
coupling parameter satisfying Robin BCs on two parallel plates in
dS spacetime. In the region between the plates, the VEVs are
decomposed into a boundary-free dS, a single plate-induced and an
interference contributions, respectively. The vacuum
energy-momentum tensor is non-diagonal, with the off-diagonal
component corresponding to the energy flux along the direction
normal to the plates. Depending on the values of the coefficients
in the boundary conditions and of the field mass this flux can be
positive or negative. In the case of a conformally coupled
massless field, the single plate contribution to the VEV of the
energy-momentum tensor vanishes and the interference part is
obtained from the corresponding result for the Minkowski bulk, by
standard conformal transformation.

The vacuum forces acting on the plates are determined by the $_{D}^{D}$%
-component of the stress. The normal stresses on the plates are presented as
sums of single plate and interaction contributions. The contributions to the
vacuum force coming from the single plate terms are the same from the left
and from the right sides of the plate and thus give no contribution to the
effective force. The interaction forces per unit surface are determined by
formula (\ref{pintj}) for general Robin BCs and by Eqs.~(\ref{pjD}),(\ref%
{pjN}) in the special cases of Dirichlet and Neumann BCs. At small distances
between the plates the vacuum forces are attractive, except for the case of
Dirichlet BC on one plate and non-Dirichlet on the other, in which case the
force turns out to be repulsive. At large separations and for positive
values of $\nu $, the force acting on the plate decays monotonically as $%
1/(2a/\eta )^{D-2\nu +2}$, for non-Neumann BCs, and as $1/(2a/\eta )^{D-2\nu
}$, in the case of Neumann BCs [see Eqs.~(\ref{pjlarge})]. For imaginary
values of $\nu $ the behavior of the vacuum forces is damping oscillatory,
in the leading order described by Eqs.~(\ref{pjlargeim}).

From the analysis carried out above, it follows that the curvature of the
background spacetime decisively influences the behavior of the Casimir
forces at distances larger than the curvature scale. As we have seen, in dS
spacetime the decay of the forces at large separations between the plates is
power-law. This is quite remarkable and clearly in contrast with the
corresponding features of the same problem in Minkowski and AdS spacetimes.

\section*{Acknowledgments}

The author acknowledges the Organizers of the 8th Alexander
Friedmann International Seminar on Gravitation and Cosmology and
CAPES (Brazil) for a support.

\end{document}